\documentclass{article}
\usepackage{arxiv}
\usepackage[utf8]{inputenc}
\usepackage[T1]{fontenc}
\usepackage{hyperref}
\usepackage{url}
\usepackage{booktabs}
\usepackage{amsfonts}
\usepackage{amsmath}
\usepackage{amssymb}
\usepackage{nicefrac}
\usepackage{microtype}
\usepackage{cleveref}
\usepackage{graphicx}
\usepackage{natbib}
\usepackage{doi}
\usepackage{bm}
\usepackage{subcaption}
\usepackage{tikz}
\usetikzlibrary{decorations.pathmorphing, arrows.meta}
\usepackage{epstopdf}

\graphicspath{{figures/}}

\title{Rapid Worst-Case Gust Identification for Very Flexible Aircraft Using Reduced-Order Models}

\author{
  Nikolaos D.~Tantaroudas\thanks{Corresponding author. Senior Researcher, ICCS.} \\
  Institute of Communications and Computer Systems (ICCS)\\
  9 Iroon Politechniou Street, Zografou, Athens 15773, Greece \\
  \texttt{nikolaos.tantaroudas@iccs.gr} \\
  \And
  Ilias Karachalios \\
  National Technical University of Athens\\
  Zografou, Athens 15780, Greece \\
}

\renewcommand{\shorttitle}{Worst-Case Gust Identification Using ROMs}

\hypersetup{
  pdftitle={Rapid Worst-Case Gust Identification for Very Flexible Aircraft Using Reduced-Order Models},
  pdfsubject={physics.flu-dyn, cs.CE},
  pdfauthor={N.D. Tantaroudas, I. Karachalios},
  pdfkeywords={Worst-case gust, Flexible aircraft, Reduced-order model, Gust loads, Certification}
}

\begin{document}
\maketitle

\begin{abstract}
Identification of worst-case gust loads is a critical step in the certification of very flexible aircraft, yet the computational cost of nonlinear full-order simulations renders exhaustive parametric searches impractical. This paper presents a reduced-order model (ROM) based methodology for rapid worst-case gust identification that achieves computational speedups of up to 600 times relative to full-order nonlinear simulations. The approach employs nonlinear model order reduction via Taylor series expansion and eigenvector projection of the coupled fluid-structure-flight dynamic system. Three test cases of increasing complexity are considered: a three-degree-of-freedom aerofoil (14 states, worst-case identified from 1,000 design sites), a Global Hawk-like UAV (540 states, 80 parametric calculations with 30$\times$ speedup), and a very flexible flying-wing (1,616 states, 37 parametric calculations reduced from 222 hours to 22 minutes). The linear ROM is shown to be accurate for deformations below 10\% of the wingspan, while the nonlinear ROM with second-order Taylor expansion accurately captures the large-deformation regime. The methodology provides a practical tool for integrating worst-case gust search into aircraft certification workflows.
\end{abstract}


\section*{Nomenclature}
\label{sec:nomenclature}

\noindent\textit{Latin}

\begin{tabbing}
12345678901234 \= \kill
$a$            \> = non-dimensional elastic axis position from mid-chord \\
$b$            \> = semichord \\
$c_h$          \> = non-dimensional distance from mid-chord to flap hinge \\
$D_{kij}$      \> = bilinear interaction coefficients of the ROM \\
$f_g$          \> = gust frequency, $U_\infty / H_g$ \\
$H_g$          \> = gust gradient distance \\
$I_\alpha$     \> = second moment of inertia of aerofoil about elastic axis \\
$I_\delta$     \> = flap moment of inertia \\
$J$            \> = structural response metric \\
$K_\xi,\,K_\alpha,\,K_\delta$ \> = plunge, torsional, and flap stiffness \\
$K_{\xi 3},\,K_{\alpha 3}$    \> = cubic nonlinear stiffness terms in plunge and pitch \\
$L_w$          \> = turbulence scale length \\
$m$            \> = number of retained ROM modes (or aerofoil sectional mass) \\
$n$            \> = number of full-order degrees of freedom \\
$r_a$          \> = radius of gyration about elastic axis, $\sqrt{I_\alpha / m\,b^2}$ \\
$\bm{R}$       \> = residual vector \\
$t$            \> = physical time \\
$U_\infty$     \> = freestream velocity \\
$U^*$          \> = reduced velocity, $U_\infty / (b\,\omega_\alpha)$ \\
$U_L^*$        \> = linear flutter reduced velocity \\
$\bm{w}$       \> = full-order state vector \\
$w_0$          \> = peak (design) gust velocity \\
$w_g$          \> = gust vertical velocity \\
$x_\alpha$     \> = aerofoil static unbalance, $S_\alpha / (m\,b)$ \\
$x_\delta$     \> = reduced centre-of-gravity distance from flap hinge \\
$\mathbf{z}$   \> = reduced-order state vector
\end{tabbing}

\noindent\textit{Greek}

\begin{tabbing}
12345678901234 \= \kill
$\alpha$       \> = angle of attack (pitch) \\
$\alpha_0$     \> = trim angle of attack \\
$\delta$       \> = trailing-edge flap deflection \\
$\lambda_k$    \> = $k$-th eigenvalue of the Jacobian \\
$\Lambda$      \> = wing sweep angle \\
$\mu$          \> = mass ratio, $m / (\pi\,\rho_\infty\,b^2)$ \\
$\xi$          \> = non-dimensional plunge displacement, $h/b$ \\
$\Omega$       \> = spatial frequency \\
$\omega_\xi,\,\omega_\alpha,\,\omega_\delta$ \> = uncoupled plunge, pitch, and flap natural frequencies \\
$\boldsymbol{\Phi},\,\boldsymbol{\Psi}$ \> = right and left eigenvector matrices \\
$\rho_\infty$  \> = freestream air density \\
$\sigma_w$     \> = turbulence intensity (RMS gust velocity) \\
$\tau$         \> = non-dimensional time, $t\,U_\infty / b$
\end{tabbing}

\noindent\textit{Acronyms}

\begin{tabbing}
12345678901234 \= \kill
DOF  \> = degree(s) of freedom \\
HALE \> = high-altitude long-endurance \\
MAC  \> = mean aerodynamic chord \\
MOR  \> = model order reduction \\
NFOM \> = nonlinear full-order model \\
NMOR \> = nonlinear model order reduction \\
POD  \> = proper orthogonal decomposition \\
PSD  \> = power spectral density \\
ROM  \> = reduced-order model \\
UAV  \> = unmanned aerial vehicle \\
UVLM \> = unsteady vortex-lattice method \\
VFA  \> = very flexible aircraft
\end{tabbing}

\section{Introduction}
\label{sec:intro}

Airworthiness certification standards CS-25 and FAR-25 require the identification of worst-case gust loads across a range of flight conditions, gust lengths, and intensities. For conventional aircraft with moderate wing flexibility, this process relies on linear analysis tools that can evaluate thousands of gust parameter combinations with modest computational effort. However, next-generation high-altitude long-endurance (HALE) and solar-powered aircraft feature very high-aspect-ratio wings that undergo large structural deformations during flight, invalidating the linear superposition principle and requiring fully nonlinear coupled aeroelastic-flight dynamic simulations~\citep{Patil2001, Noll2004, Patil2006}.

The computational cost of nonlinear simulations presents a severe bottleneck for certification. A single time-domain simulation of the coupled system, with hundreds to thousands of degrees of freedom (DOF), typically requires several hours of wall-clock time on a single processor. A worst-case gust search involving $\mathcal{O}(10^2)$ to $\mathcal{O}(10^3)$ parameter combinations therefore demands weeks to months of computation, rendering exhaustive searches impractical during the design cycle. The NASA Helios mishap investigation~\citep{Noll2004} highlighted the critical importance of understanding gust response for very flexible aircraft, as turbulence-induced large deformations contributed to the catastrophic failure.

Model order reduction (MOR) offers a systematic approach to overcoming this computational barrier. Various techniques have been explored for aeroelastic applications, including proper orthogonal decomposition (POD)~\citep{Lucia2004}, harmonic balance~\citep{Badcock2011}, and system identification methods~\citep{Dowell2001}. However, these approaches either require excitation-specific training data (POD), are restricted to periodic regimes (harmonic balance), or cannot capture essential geometric nonlinearities (linear identification). The nonlinear model order reduction (NMOR) technique employed in this work~\citep{DaRonch2013gust, DaRonch2013control, Tantaroudas2017bookchapter} projects the coupled system onto a compact eigenvector basis of the Jacobian matrix, retaining nonlinear terms through a Taylor series expansion. Once constructed, the ROM is excitation-independent and can be applied to arbitrary gust profiles without regeneration.

Since 2015, significant advances have been made in reduced-order modelling for flexible aircraft. Wang et al.~\citep{Wang2015compstruc} developed a normal-mode-based model reduction technique for geometrically nonlinear slender structures using intrinsic beam equations, which was subsequently extended to a full nonlinear modal aeroservoelastic framework for flexible aircraft~\citep{Wang2016aiaa}. These modal approaches were further refined by Artola et al.~\citep{Artola2021}, who demonstrated aeroelastic control and estimation using a minimal nonlinear modal description with moving-horizon estimation and model-predictive control, while Goizueta et al.~\citep{Goizueta2022} introduced adaptive sampling strategies for interpolation of parametric ROMs across flight conditions. Riso and Cesnik~\citep{Riso2023} systematically assessed the impact of low-order modelling on aeroelastic predictions for very flexible wings, confirming the suitability of beam-based ROM approaches for large-deformation regimes. On the worst-case gust identification front, Balatti et al.~\citep{Balatti2022} investigated the effect of folding wingtips on worst-case gust loads, illustrating the continuing need for efficient parametric search methodologies. A comprehensive treatment of the coupled flight mechanics, aeroelasticity, and control of flexible aircraft, including gust response modelling, is provided in the recent monograph by Palacios and Cesnik~\citep{PalaciosCesnik2023}.

The NMOR approach was originally developed for gust loads analysis in~\citep{DaRonch2013gust} and for nonlinear control applications in~\citep{DaRonch2013control}. It was subsequently extended to free-flying flexible aircraft with coupled flight dynamics in~\citep{Tantaroudas2015scitech, Tantaroudas2014aviation}, validated experimentally against wind tunnel data in~\citep{DaRonch2014flutter, Papatheou2013ifasd, Fichera2014isma}, and comprehensively documented in the book chapter~\citep{Tantaroudas2017bookchapter}. The impact of aerodynamic modelling fidelity on the framework was assessed in~\citep{DaRonch2014scitech_flight}. A detailed and self-contained derivation of the NMOR formulation for coupled aeroelastic-flight dynamic systems, including third-order Taylor expansion terms and systematic eigenvector selection criteria, is presented in the companion paper~\citep{Tantaroudas2026nmor}.

The contributions of this paper are: (i) a systematic demonstration of the ROM-based worst-case gust search methodology across three test cases of increasing complexity; (ii) quantification of achievable computational speedups (up to 600$\times$); (iii) delineation of the validity domains of linear versus nonlinear ROMs; and (iv) practical guidance for integration into aircraft certification workflows.

\section{Gust Modelling}
\label{sec:gust}

\subsection{Discrete Gust: 1-Minus-Cosine Profile}

\begin{figure}[htbp]
\centering
\includegraphics[width=0.55\textwidth]{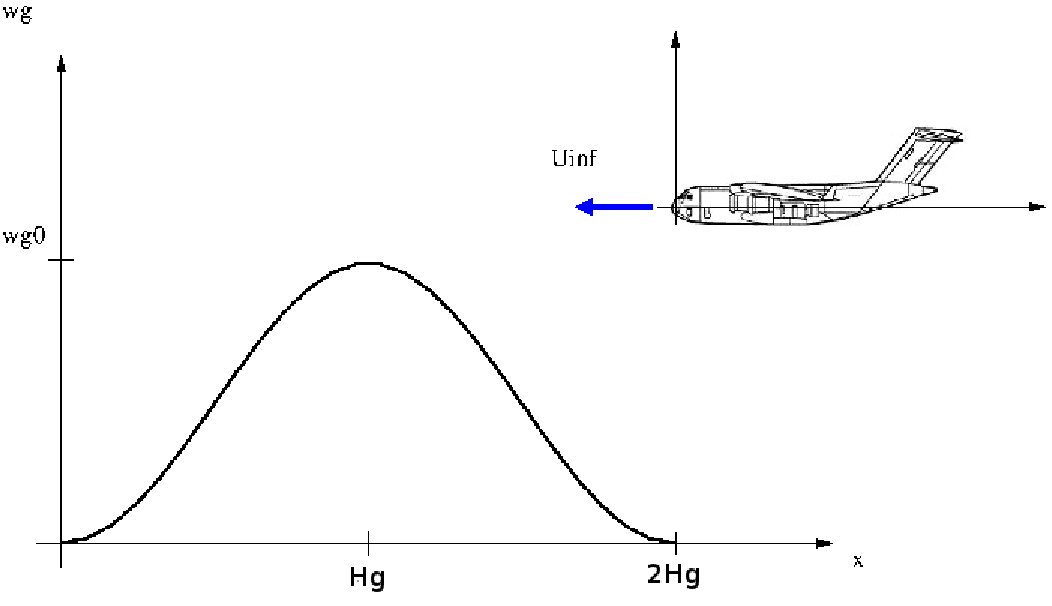}
\caption{The ``1-minus-cosine'' discrete gust profile prescribed by CS-25/FAR-25 certification standards. The gust is parameterised by the peak velocity $w_0$ and the gradient distance $H_g$. The worst-case identification requires searching over the space of $H_g$ values at each flight condition.}
\label{fig:oneminus_cosine}
\end{figure}

The standard discrete gust profile prescribed by CS-25.341(a) and FAR-25.341(a) is the ``1-minus-cosine'' gust, illustrated in~\Cref{fig:oneminus_cosine}:
\begin{equation}
w_g(x) = \frac{w_0}{2}\left(1 - \cos\left(\frac{2\pi x}{H_g}\right)\right), \qquad 0 \leq x \leq H_g
\label{eq:1mc}
\end{equation}
where $w_0$ is the peak gust velocity (design gust velocity), $H_g$ is the gust gradient distance, and $x = U_\infty t$ is the distance penetrated into the gust at the aircraft forward speed $U_\infty$. The gust velocity is zero outside the interval $[0, H_g]$.

The certification standards specify that $H_g$ must be varied over a range of gradient distances, with the design gust velocity $w_0$ prescribed as a function of altitude and gust gradient distance. The worst-case identification problem therefore reduces to finding the critical $H_g$ that maximises the structural response metric of interest.

For a time-domain simulation, the spatial gust profile is converted to a temporal profile using the frozen-turbulence (Taylor) hypothesis:
\begin{equation}
w_g(t) = \frac{w_0}{2}\left(1 - \cos\left(\frac{2\pi U_\infty (t - t_0)}{H_g}\right)\right), \qquad t_0 \leq t \leq t_0 + \frac{H_g}{U_\infty}
\label{eq:1mc_time}
\end{equation}
where $t_0$ is the gust onset time. The gust frequency content is centred at $f_g = U_\infty / H_g$; short gusts have high-frequency content that excites structural modes, while long gusts excite rigid-body flight dynamics.

\subsection{Continuous Turbulence: Von K\'arm\'an Spectrum}

For continuous turbulence analysis as required by CS-25.341(b), the Von K\'arm\'an power spectral density (PSD) of vertical gust velocity is employed:
\begin{equation}
\Phi_{ww}(\Omega) = \sigma_w^2 \frac{L_w}{\pi} \frac{1 + \frac{8}{3}\left(1.339\,L_w \Omega\right)^2}{\left(1 + \left(1.339\,L_w\Omega\right)^2\right)^{11/6}}
\label{eq:vk_psd}
\end{equation}
where $\sigma_w$ is the turbulence intensity (root-mean-square gust velocity), $L_w$ is the turbulence scale length (typically 2,500~ft at high altitude), and $\Omega = 2\pi f / U_\infty$ is the spatial frequency. Time-domain realisations of Von K\'arm\'an turbulence are generated by passing white noise through a rational approximation filter matched to~\Cref{eq:vk_psd}~\citep{Dowell2004}.

\subsection{Gust Penetration Effects}

For aircraft with long wing spans, gust penetration, the time delay between the gust encountering the wing root and reaching the wing tip, introduces spanwise phase variations in the gust loading. Each spanwise strip $i$ at location $y_i$ experiences the gust with a time delay:
\begin{equation}
\Delta t_i = \frac{y_i \sin\Lambda}{U_\infty}
\label{eq:gust_penetration}
\end{equation}
where $\Lambda$ is the wing sweep angle. For the straight-wing configurations considered here ($\Lambda = 0$), gust penetration reduces to a uniform encounter along the span.

\section{Aeroelastic Model and Reduction}
\label{sec:model}

\subsection{Coupled System Formulation}

The coupled aeroelastic-flight dynamic system is expressed in first-order state-space form~\citep{Tantaroudas2017bookchapter}:
\begin{equation}
\frac{d\mathbf{w}}{dt} = \mathbf{R}(\mathbf{w}, \mathbf{u}_c, \mathbf{u}_d)
\end{equation}
with state vector $\mathbf{w} = \{\mathbf{w}_f, \mathbf{w}_s, \mathbf{w}_r\}^T$ partitioned into aerodynamic ($\mathbf{w}_f$), structural ($\mathbf{w}_s$), and rigid-body ($\mathbf{w}_r$) states. The aerodynamic model uses unsteady strip theory with Wagner and K\"ussner indicial functions~\citep{Theodorsen1935, Wagner1925, Kussner1936}, providing time-domain aerodynamic loads through augmented states. The structural model employs geometrically-exact nonlinear beam elements~\citep{Hodges2003, Palacios2010} with six DOF per node, and the flight dynamics use quaternion-based attitude propagation~\citep{Tantaroudas2015scitech}.

The gust enters the system through the disturbance input:
\begin{equation}
\mathbf{u}_d(t) = \left\{ w_g(t, y_1), \; w_g(t, y_2), \; \ldots, \; w_g(t, y_{N_s}) \right\}^T
\end{equation}
where $w_g(t, y_i)$ is the gust velocity at strip $i$, accounting for gust penetration delays. The K\"ussner function models the unsteady lift build-up due to gust entry at each strip, introducing two augmented aerodynamic states per strip for the gust coupling.

\subsection{Nonlinear Model Order Reduction}

The NMOR procedure follows the Taylor-expansion and eigenvector-projection approach detailed in~\citep{DaRonch2013control, DaRonch2013gust, Tantaroudas2015scitech}. The nonlinear residual is expanded around the trimmed equilibrium $\mathbf{w}_0$ up to second order:
\begin{equation}
\mathbf{R}(\mathbf{w}) \approx \mathbf{A}\Delta\mathbf{w} + \frac{1}{2}\mathcal{B}(\Delta\mathbf{w}, \Delta\mathbf{w}) + \mathbf{B}_g \Delta\mathbf{u}_d
\label{eq:taylor}
\end{equation}
where $\mathbf{A}$ is the Jacobian at equilibrium, $\mathcal{B}$ is the symmetric bilinear operator encoding second-order nonlinear interactions, and $\mathbf{B}_g$ is the gust input matrix. The system is projected onto $m$ biorthonormal eigenvectors of the Jacobian:
\begin{equation}
\Delta\mathbf{w} \approx \boldsymbol{\Phi}\mathbf{z} + \bar{\boldsymbol{\Phi}}\bar{\mathbf{z}}, \qquad \boldsymbol{\Psi}^H \boldsymbol{\Phi} = \mathbf{I}_m
\end{equation}
yielding the reduced-order equations:
\begin{equation}
\dot{z}_k = \lambda_k z_k + \sum_{i,j} D_{kij} z_i z_j + \boldsymbol{\psi}_k^H \mathbf{B}_g \Delta\mathbf{u}_d, \qquad k = 1, \ldots, m
\label{eq:rom}
\end{equation}
The bilinear interaction coefficients $D_{kij}$ are computed via matrix-free finite differences~\citep{DaRonch2013control}, requiring $\mathcal{O}(m^2)$ residual evaluations. The eigenvector basis is selected systematically: real eigenvalues near the origin capture rigid-body and gust coupling dynamics, while lightly damped complex eigenvalues capture structural bending and torsion modes~\citep{Tantaroudas2015scitech}.

A critical property of the ROM for worst-case gust searches is that it is \emph{excitation-independent}: the coefficients $\lambda_k$ and $D_{kij}$ are computed once during ROM construction, and the ROM can subsequently be excited by any gust profile (different $H_g$, $w_0$, Von K\'arm\'an realisations) without regeneration. This makes parametric sweeps over gust parameters computationally trivial.

\section{Worst-Case Gust Search Methodology}
\label{sec:methodology}

The worst-case gust search is formulated as a constrained optimisation problem:
\begin{equation}
H_g^* = \arg\max_{H_g \in [H_{\min}, H_{\max}]} \; J\!\left(\mathbf{w}(t; H_g, w_0(H_g))\right)
\label{eq:wcgs_opt}
\end{equation}
where $J$ is the structural response metric, typically the maximum wing-tip vertical displacement $\max_t |w_{\text{tip}}(t)|$ or the maximum root bending moment $\max_t |M_{\text{root}}(t)|$, and $w_0(H_g)$ is the design gust velocity prescribed by the certification standard as a function of gust gradient distance.

\begin{figure}[htbp]
\centering
\includegraphics[width=0.55\textwidth]{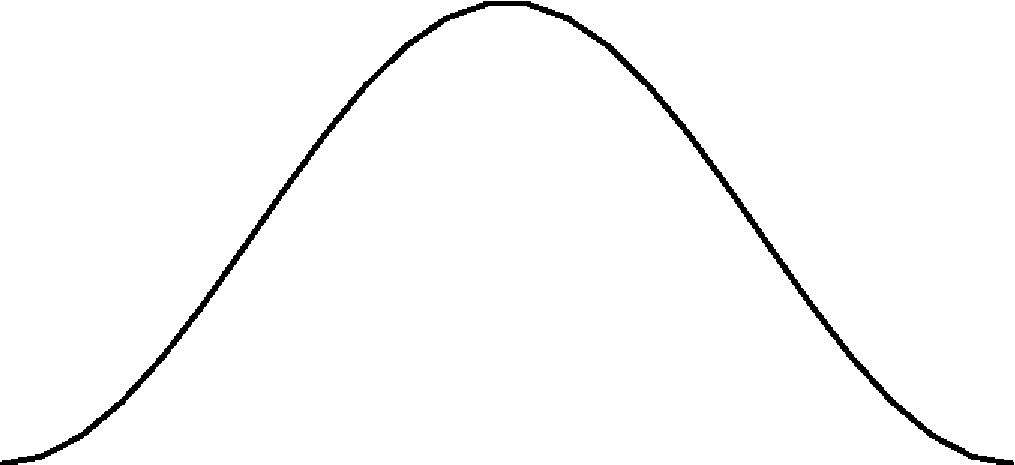}
\caption{Schematic of gust encounter: the aircraft penetrates a ``1-minus-cosine'' gust of gradient distance $H_g$ at freestream velocity $U_\infty$. The gust velocity $w_g$ acts as a vertical perturbation to the aerodynamic angle of attack at each spanwise strip.}
\label{fig:gust_profile}
\end{figure}

The gust profile enters the simulation as shown in~\Cref{fig:gust_profile}. The ROM-based search strategy proceeds as follows:
\begin{enumerate}
\item \textbf{ROM construction} (one-time cost): compute the Jacobian eigenspectrum at the trim state, select the $m$-mode basis, and evaluate the bilinear interaction coefficients $D_{kij}$.
\item \textbf{Parametric sweep}: for each candidate $H_g$, time-march the $m$-dimensional ROM~\eqref{eq:rom} and record the response metric $J$.
\item \textbf{Worst-case identification}: determine $H_g^*$ from the ROM-based sweep.
\item \textbf{Validation}: confirm the identified worst-case with a single full-order nonlinear simulation.
\end{enumerate}

Two ROM fidelity levels are compared. The \emph{linear ROM} retains only the Jacobian term ($\dot{z}_k = \lambda_k z_k + \boldsymbol{\psi}_k^H \mathbf{B}_g \Delta\mathbf{u}_d$) and is valid for small perturbations where geometric nonlinearities are negligible (tip deflections below approximately 10\% of the wingspan). The \emph{nonlinear ROM} additionally retains the second-order Taylor expansion terms ($D_{kij}$), capturing geometric stiffening and large-deformation coupling effects essential for very flexible configurations.

\section{Results}
\label{sec:results}

\subsection{Three-Degree-of-Freedom Aerofoil}
\label{sec:3dof}

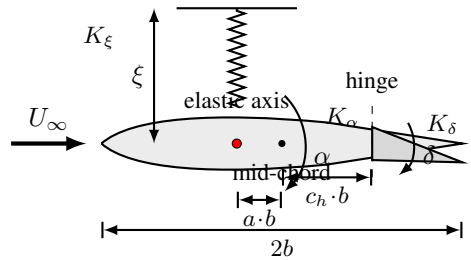
\begin{figure}[htbp]
\centering
\begin{tikzpicture}[scale=1.0, >=latex]
  \draw[thick, fill=gray!15]
    (0,0) .. controls (0.4,0.35) and (1.5,0.42) .. (3.0,0.28)
    -- (4.8,0.0)
    -- (3.0,-0.28) .. controls (1.5,-0.42) and (0.4,-0.35) .. (0,0);
  \draw[dashed] (3.6,0.5) -- (3.6,-0.5);
  \draw[thick, fill=gray!30]
    (3.6,0.22) -- (4.8,-0.25) -- (3.6,-0.22) -- cycle;
  \draw[fill=black] (2.4,0) circle (0.04);
  \node[below] at (2.4,-0.15) {\small mid-chord};
  \draw[fill=red] (1.8,0) circle (0.06);
  \node[above] at (1.8,0.35) {\small elastic axis};
  \draw[thick, decorate, decoration={zigzag, segment length=4, amplitude=3}]
    (1.8,0.5) -- (1.8,1.8);
  \draw[thick] (1.0,1.8) -- (2.6,1.8); 
  \draw[thick, <->] (0.7,0.0) -- (0.7,1.8);
  \node[left] at (0.7,0.9) {$\xi$};
  \node[left] at (0.3,1.4) {\small $K_\xi$};
  \draw[thick, ->] (1.8,0) ++(45:0.9) arc (45:-45:0.9);
  \node[right] at (2.7,-0.2) {$\alpha$};
  \node[right] at (2.85,0.35) {\small $K_\alpha$};
  \draw[thick, ->] (3.6,0) ++(30:0.55) arc (30:-50:0.55);
  \node[right] at (4.15,-0.15) {$\delta$};
  \node[right] at (4.2,0.25) {\small $K_\delta$};
  \draw[thick, |<->|] (1.8,-0.75) -- (2.4,-0.75);
  \node[below] at (2.1,-0.75) {\small $a\!\cdot\!b$};
  \draw[thick, |<->|] (0,-1.15) -- (4.8,-1.15);
  \node[below] at (2.4,-1.15) {\small $2b$};
  \draw[ultra thick, ->] (-1.2,0) -- (-0.2,0);
  \node[above] at (-0.7,0.05) {$U_\infty$};
  \node[above] at (3.6,0.55) {\small hinge};
  \draw[thick, |<->|] (2.4,-0.45) -- (3.6,-0.45);
  \node[below] at (3.0,-0.45) {\small $c_h\!\cdot\!b$};
\end{tikzpicture}
\caption{Schematic of the three-DOF aerofoil test case showing the pitch ($\alpha$), plunge ($\xi$), and trailing-edge flap ($\delta$) degrees of freedom. The elastic axis is located at distance $a \cdot b$ from mid-chord; the flap hinge at $c_h \cdot b$ from mid-chord. Springs $K_\alpha$, $K_\xi$, and $K_\delta$ represent the restoring stiffness in each DOF.}
\label{fig:patil_schematic}
\end{figure}

The first test case is a two-dimensional aerofoil with three DOF (pitch $\alpha$, plunge $\xi$, trailing-edge flap $\delta$) and 14 full-order states (8 aerodynamic + 6 structural), reduced to a 4-mode ROM~\citep{DaRonch2013gust, Fichera2014isma}. The aerofoil schematic is shown in~\Cref{fig:patil_schematic}. Cubic hardening nonlinearities are present in pitch and plunge stiffness: $K_{\alpha_3} = 3$, $K_{\xi_3} = 1.0$.

The worst-case gust search spans 1,000 gust gradient distances in the range $H_g \in [0.1, 100]$ semichords at freestream velocity $U^* = 4.5$ (below the linear flutter speed $U_L^* = 6.37$). The peak gust velocity is set to 14\% of the freestream. The response metric is the maximum plunge displacement.

\begin{figure}[htbp]
\centering
\includegraphics[width=0.6\textwidth]{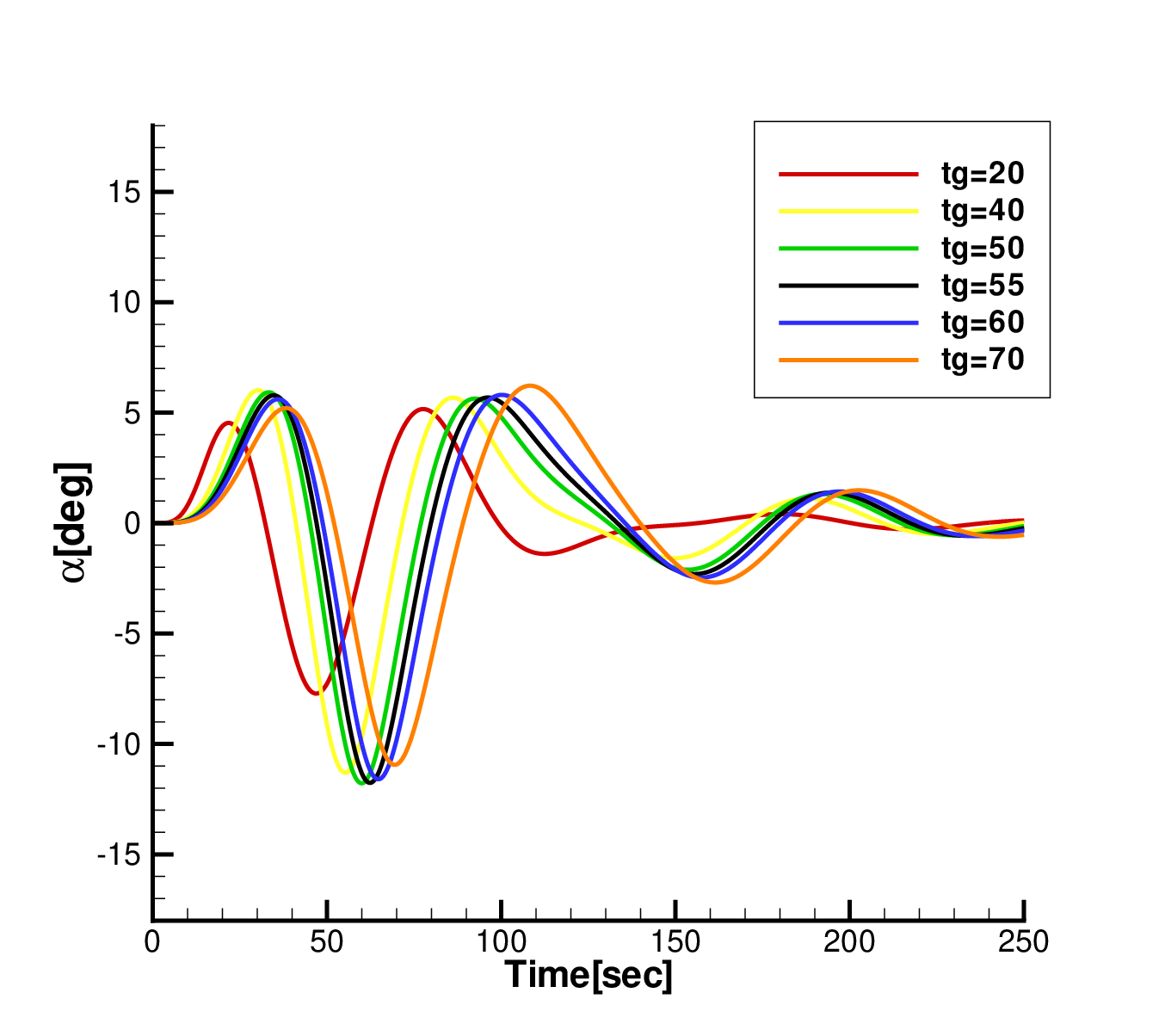}
\caption{Time-domain pitch angle responses of the three-DOF aerofoil at selected gust gradient distances $t_g$ (in semichords). The response at $t_g = 55$ semichords (bold black) exhibits the largest oscillation amplitude, corresponding to resonant excitation near the plunge natural frequency.}
\label{fig:worst_case_3dof}
\end{figure}

\Cref{fig:worst_case_3dof} shows representative time-domain pitch angle responses at selected gust gradient distances. The response amplitude varies significantly with the gust duration, exhibiting the largest oscillation at $t_g \approx 55$ semichords, which corresponds to a gust frequency in the vicinity of the plunge natural frequency. Notably, the worst-case is not at the longest gust (which would produce the largest quasi-static load) but at an intermediate length where frequency-content resonance amplifies the dynamic response.

The nonlinear ROM reproduces the full-order response across the parameter space. The ROM enables evaluation of all 1,000 design sites at a fraction of the computational cost required for the full-order model.

\subsection{Global Hawk-Like UAV}
\label{sec:uav}

\begin{figure}[htbp]
\centering
\includegraphics[width=0.75\textwidth]{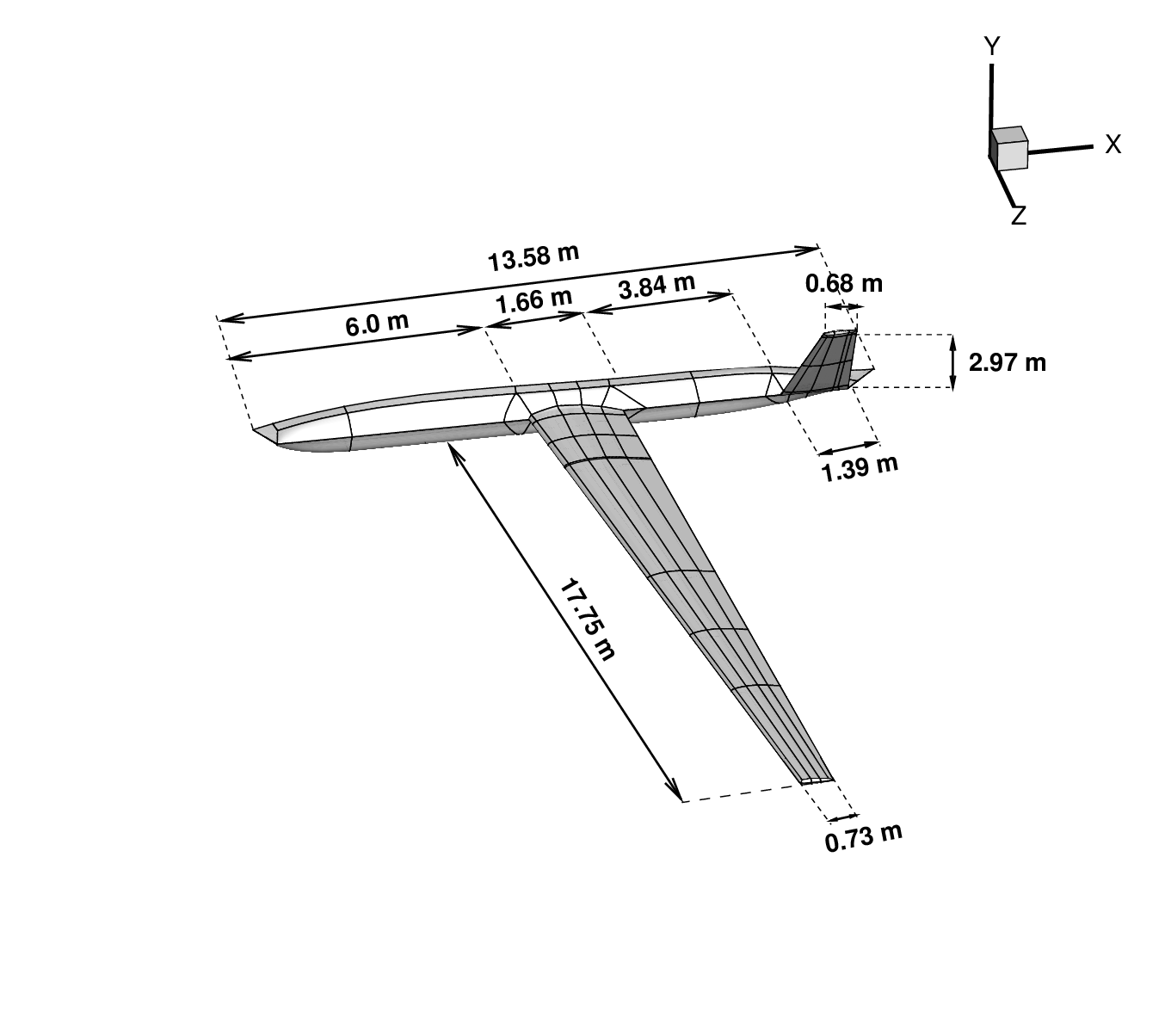}
\caption{Global Hawk-like UAV configuration used for worst-case gust search~\citep{Tantaroudas2014aviation, Tantaroudas2017bookchapter}. The model features a 17.75-m semi-span flexible wing, 13.58-m fuselage, and V-tail empennage, discretised with geometrically-exact beam elements. Key dimensions are indicated.}
\label{fig:hale_geometry}
\end{figure}

The second test case is a Global Hawk-like unmanned aerial vehicle with 17.75-m semi-span, modelled with 20 beam elements for the wing and a beam-stick representation of the fuselage and empennage~\citep{Tantaroudas2014aviation, Tantaroudas2017bookchapter}. The aircraft geometry is shown in~\Cref{fig:hale_geometry}. The full-order model has $n = 540$ DOF, reduced to an $m = 8$ mode ROM (5 structural modes + 3 gust coupling modes).

\begin{figure}[htbp]
\centering
\begin{subfigure}[b]{0.48\textwidth}
\includegraphics[width=\textwidth]{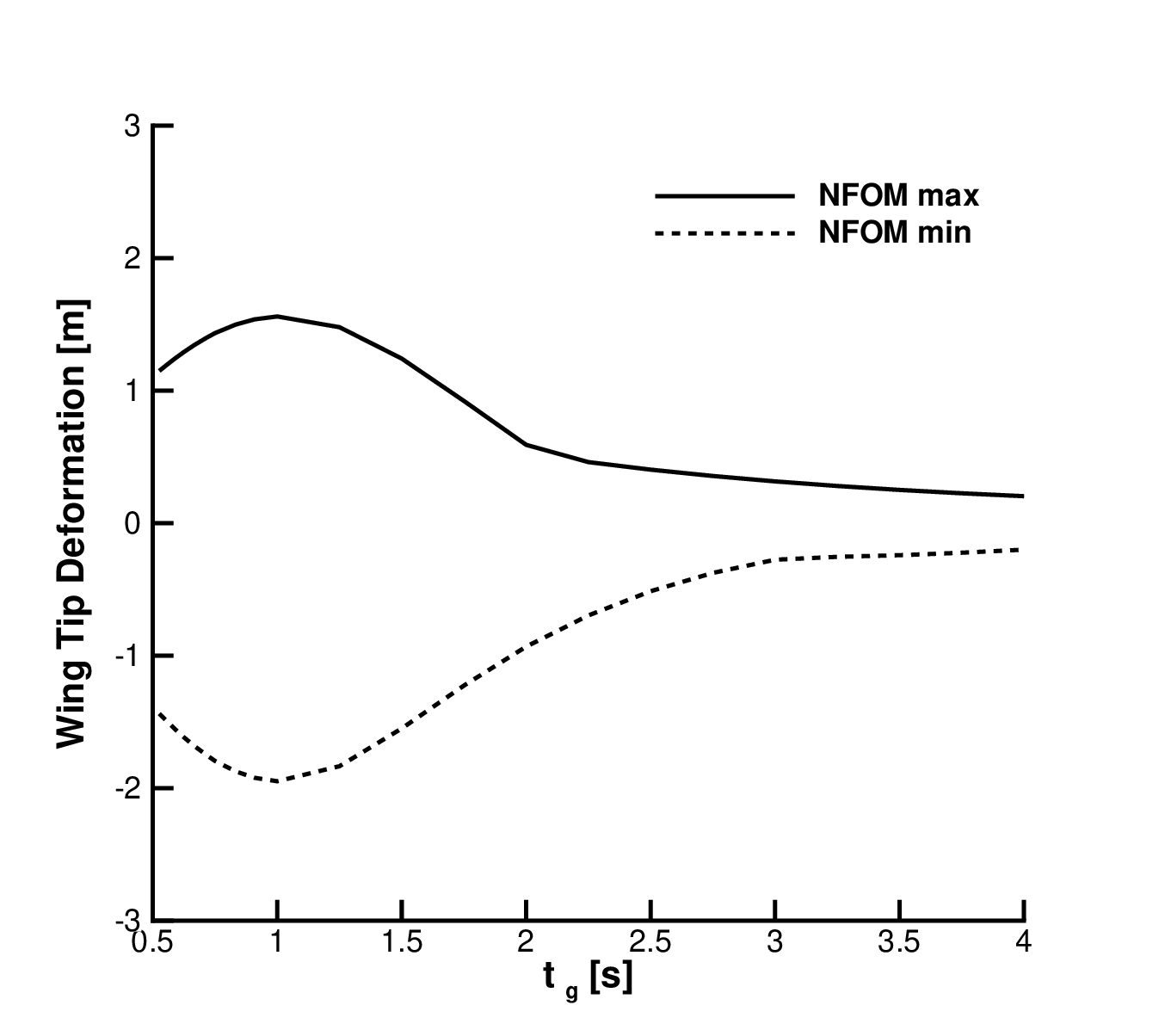}
\caption{Peak wing-tip vertical displacement}
\label{fig:hale_wcgs_wtip}
\end{subfigure}
\hfill
\begin{subfigure}[b]{0.48\textwidth}
\includegraphics[width=\textwidth]{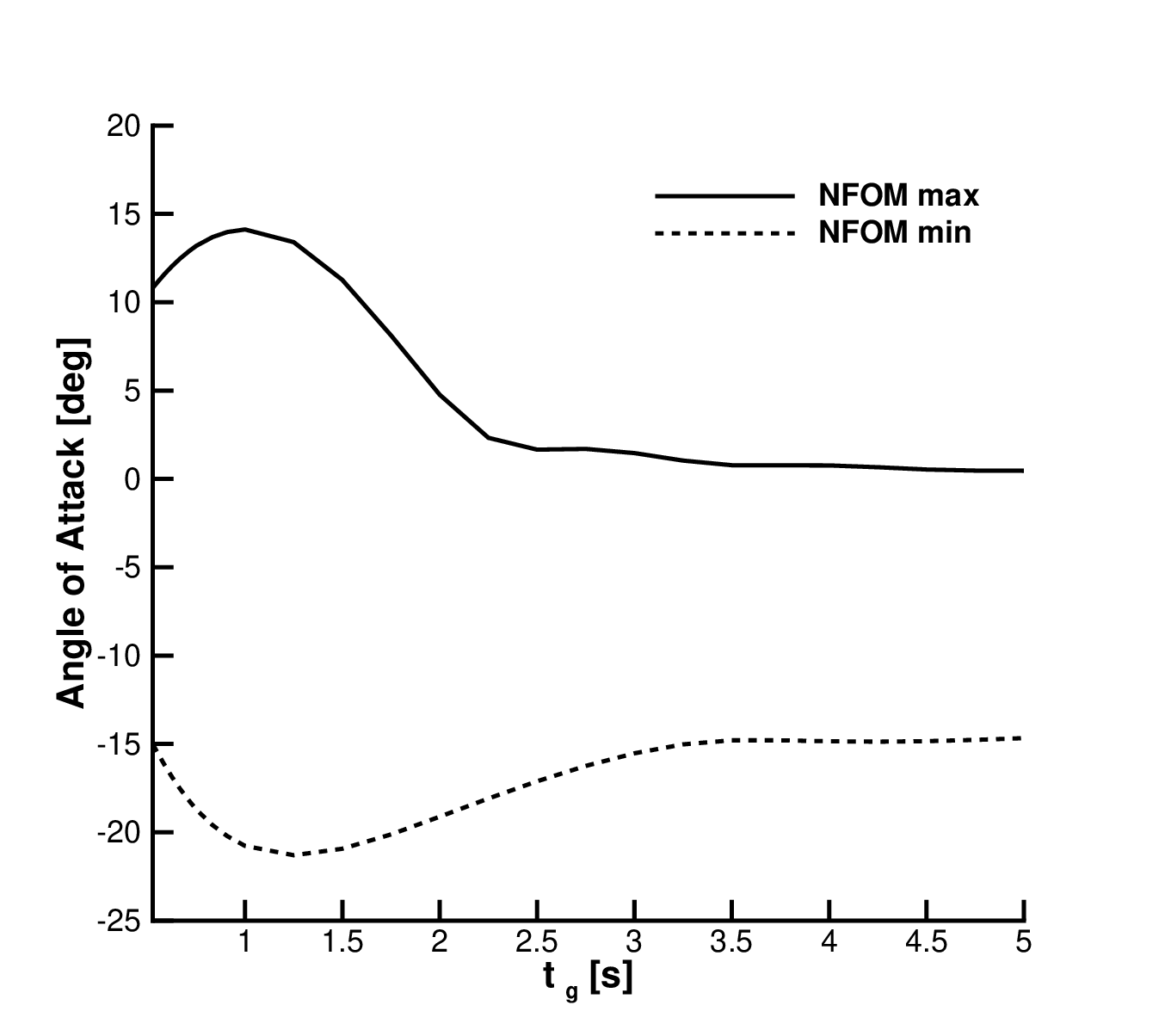}
\caption{Peak angle-of-attack perturbation}
\label{fig:hale_wcgs_aoa}
\end{subfigure}
\caption{Worst-case gust search results for the UAV configuration: peak structural response metrics as functions of gust duration. The maximum and minimum response envelopes from the full-order nonlinear model (NFOM) are shown, identifying the critical gust duration that produces the largest structural loads.}
\label{fig:hale_wcgs}
\end{figure}

Flight conditions are: $U_\infty = 59$~m/s, $\rho = 0.0789$~kg/m$^3$ (high altitude), trim angle of attack $\alpha_0 = 4^\circ$. A parametric sweep of 80 ``1-minus-cosine'' gust cases with wavelengths ranging from 0 to 776 MAC (step size 9.7) identifies the worst-case condition. The worst-case gust search results are presented in~\Cref{fig:hale_wcgs}.

\begin{table}[htbp]
\centering
\caption{Computational performance for the UAV worst-case gust search.}
\label{tab:uav_speedup}
\begin{tabular}{@{}lcc@{}}
\toprule
Model & DOFs & Approx.\ speedup \\
\midrule
Full-order nonlinear & 540 & 1$\times$ \\
Nonlinear ROM & 8 & 30$\times$ \\
\bottomrule
\end{tabular}
\end{table}

At the worst-case condition, the wing-tip deflection reaches approximately 9~m, corresponding to 50\% of the semi-span. At this deformation level, geometric stiffening is pronounced: the upward bending of the wing rotates the local lift vector inward, creating an effective dihedral that reduces the net vertical force. The linear ROM, which does not capture this effect, overpredicts the peak tip deflection. The nonlinear ROM with second-order Taylor expansion accurately predicts the saturated response~\citep{Tantaroudas2017bookchapter}.

\subsubsection{Effect of gust gradient distance}

\begin{figure}[htbp]
\centering
\includegraphics[width=0.6\textwidth]{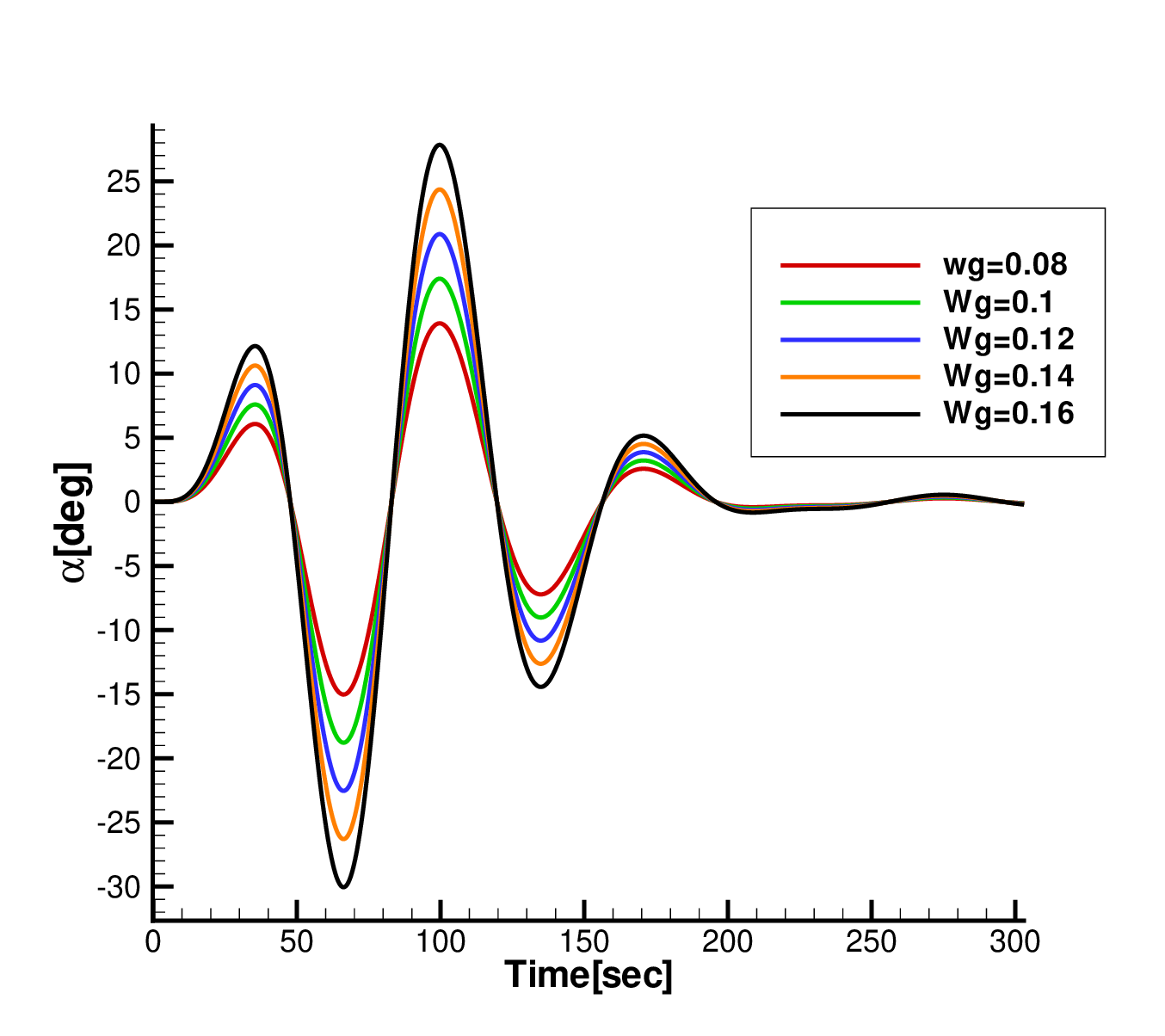}
\caption{Sensitivity of the UAV pitch angle response to gust intensity: time-domain responses for different normalised gust velocity amplitudes $w_g$. The oscillation amplitude grows with increasing gust intensity, demonstrating the importance of systematic parametric evaluation.}
\label{fig:worst_case_sweep}
\end{figure}

The sensitivity of the flight dynamic response to gust intensity is illustrated in~\Cref{fig:worst_case_sweep}, which shows the pitch angle time histories for the UAV encountering gusts of varying intensity. As the gust intensity increases, the pitch oscillation amplitude grows significantly, demonstrating the nonlinear coupling between the gust loading and the flight dynamic response. The non-monotonic dependence of the peak response on gust parameters underscores the necessity of exhaustive parametric searches rather than simplified analytical estimates.

This behaviour highlights the practical value of the ROM approach, which makes parametric sweeps computationally affordable~\citep{DaRonch2013gust, DaRonch2014scitech_flight}.

\subsection{Very Flexible Flying-Wing}
\label{sec:vfa}

The third test case is a 32-m span very flexible flying-wing with $n = 1{,}616$ full-order DOF, reduced to $m = 9$ ROM modes~\citep{Tantaroudas2015scitech}. This is an all-wing configuration without fuselage or tail, representing the most demanding scenario for gust loads analysis due to the extreme flexibility and the absence of inherent longitudinal static stability. Properties: chord 1~m, mass 10~kg/m, $EI_2 = 2.5 \times 10^4$~Nm$^2$, $GJ = 1.25 \times 10^4$~Nm$^2$. Flight conditions: altitude 20,000~m, $U_\infty = 25$~m/s, $\rho = 0.0889$~kg/m$^3$.

A parametric sweep of 37 ``1-minus-cosine'' gust cases with gust durations from 0.5~s to 20~s (corresponding to $H_g = 12.5$~m to 500~m) is performed.

\subsubsection{Worst-case identification}

\begin{figure}[htbp]
\centering
\begin{subfigure}[b]{0.48\textwidth}
\includegraphics[width=\textwidth]{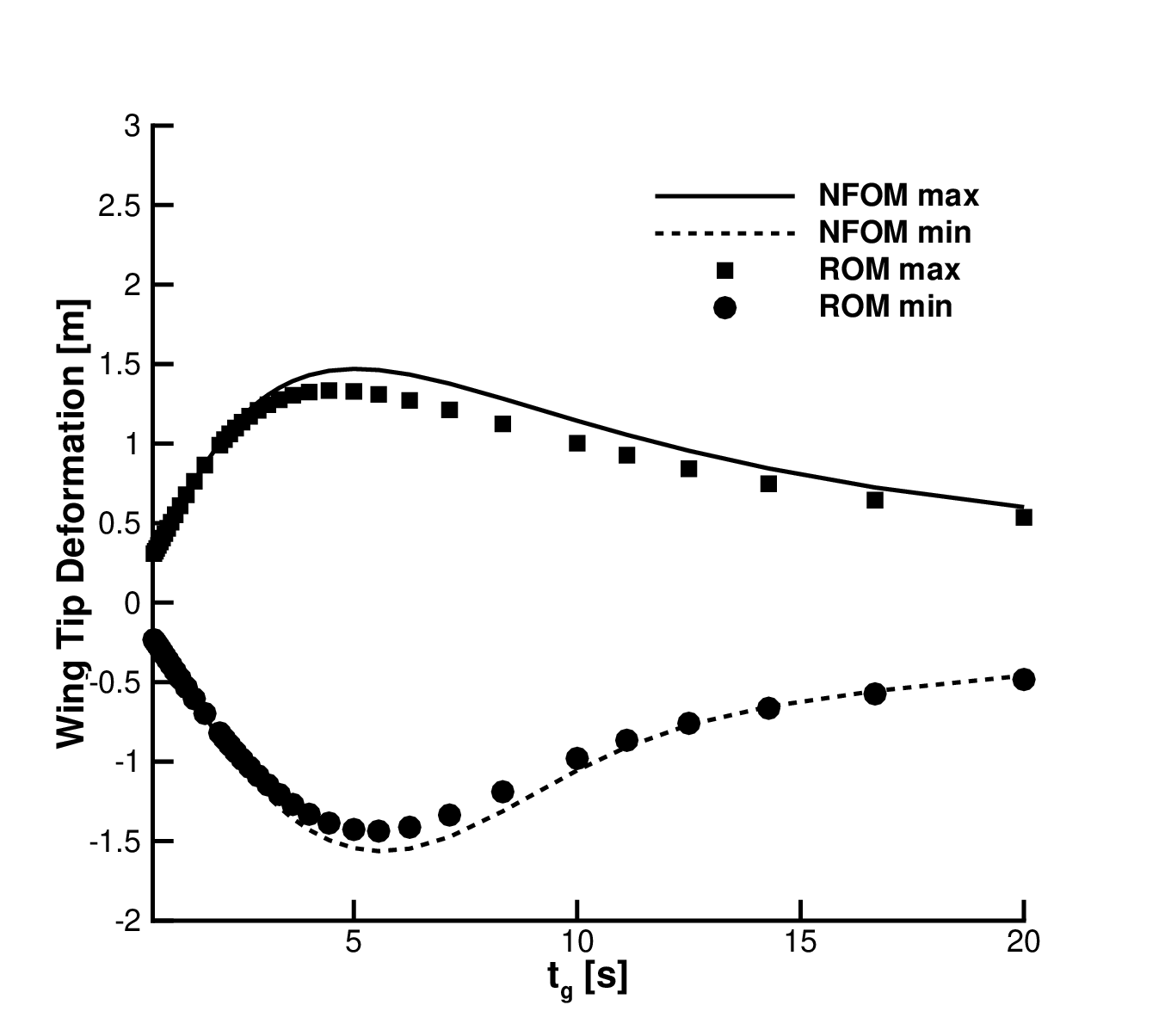}
\caption{Peak wing-tip vertical displacement}
\label{fig:vfa_wcgs_wtip}
\end{subfigure}
\hfill
\begin{subfigure}[b]{0.48\textwidth}
\includegraphics[width=\textwidth]{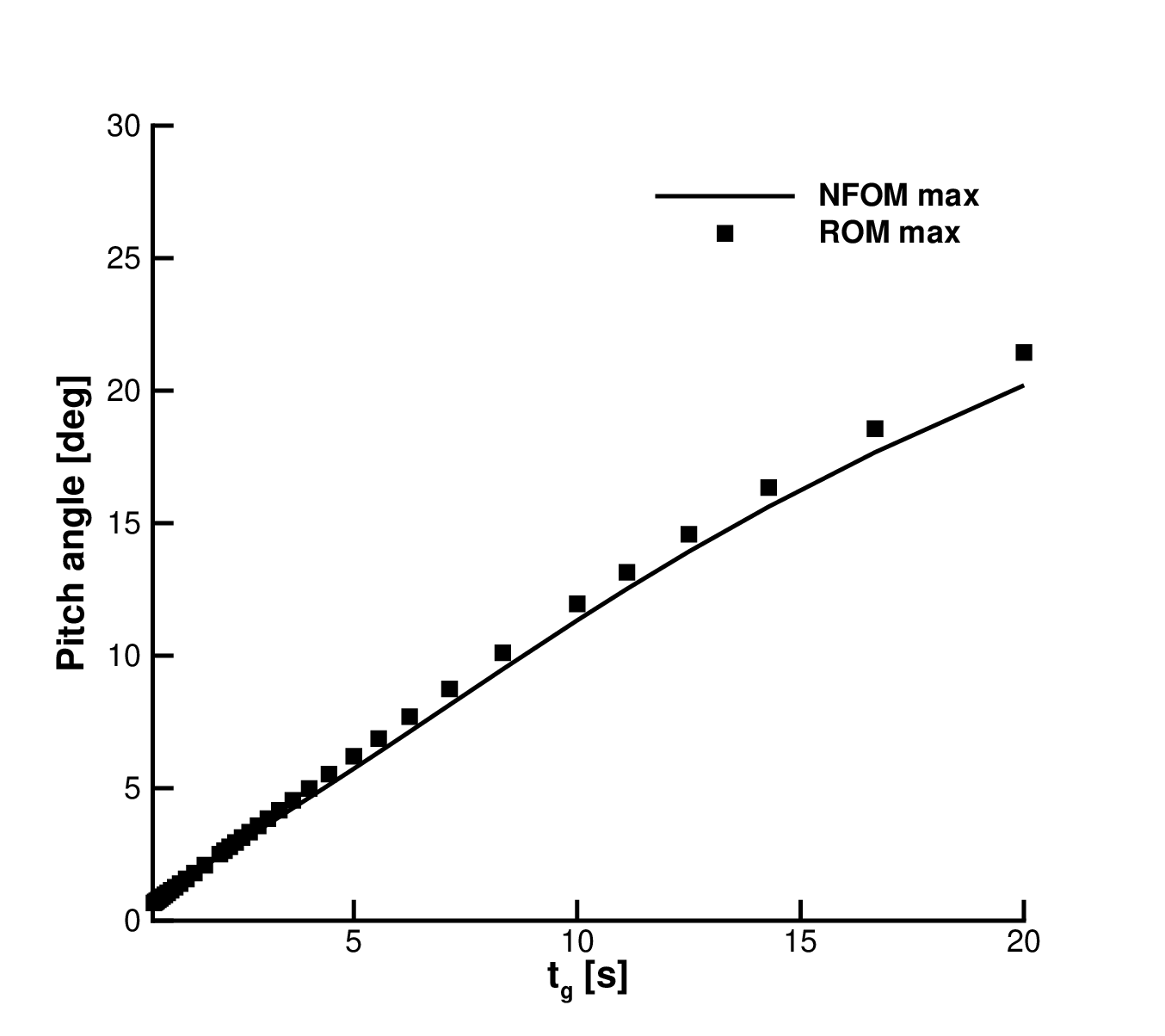}
\caption{Peak angle-of-attack perturbation}
\label{fig:vfa_wcgs_aoa}
\end{subfigure}
\caption{Worst-case gust search results for the very flexible flying-wing: peak structural response metrics as functions of gust duration. Full-order nonlinear model (circles), nonlinear ROM (solid), linear ROM (dashed). The worst-case occurs at $\sim$5.0~s gust duration.}
\label{fig:vfa_wcgs}
\end{figure}

\Cref{fig:vfa_wcgs_wtip,fig:vfa_wcgs_aoa} present the peak wing-tip vertical displacement and body-frame angle-of-attack perturbation as functions of gust duration. The worst-case gust is identified at a duration of approximately 5.0~s ($H_g = 125$~m, equivalent to 125~MAC for 1-m chord). At this condition, the maximum wing-tip deformation reaches approximately 10\% of the wingspan, with a maximum bending slope at the tip of 7.7 degrees~\citep{Tantaroudas2017bookchapter}.

\subsubsection{Computational speedup}

\begin{table}[htbp]
\centering
\caption{Computational cost comparison for the VFA worst-case gust search (37 cases).}
\label{tab:vfa_speedup}
\begin{tabular}{@{}lccc@{}}
\toprule
Model & DOFs & Total time & Speedup \\
\midrule
Full-order nonlinear & 1,616 & 222 hours & 1$\times$ \\
Nonlinear ROM & 9 & 22 minutes & \textbf{600$\times$} \\
\bottomrule
\end{tabular}
\end{table}

The 600$\times$ speedup stems from two factors: (i) the ROM time integration involves only a 9-dimensional ODE system instead of a 1,616-dimensional one, with correspondingly smaller matrices; and (ii) the ROM does not require Newton-type iterations for nonlinear equilibrium at each time step, as the nonlinear terms are explicit quadratic functions of the reduced states.

\subsubsection{Linear versus nonlinear ROM accuracy}

\begin{figure}[htbp]
\centering
\begin{subfigure}[b]{0.48\textwidth}
\includegraphics[width=\textwidth]{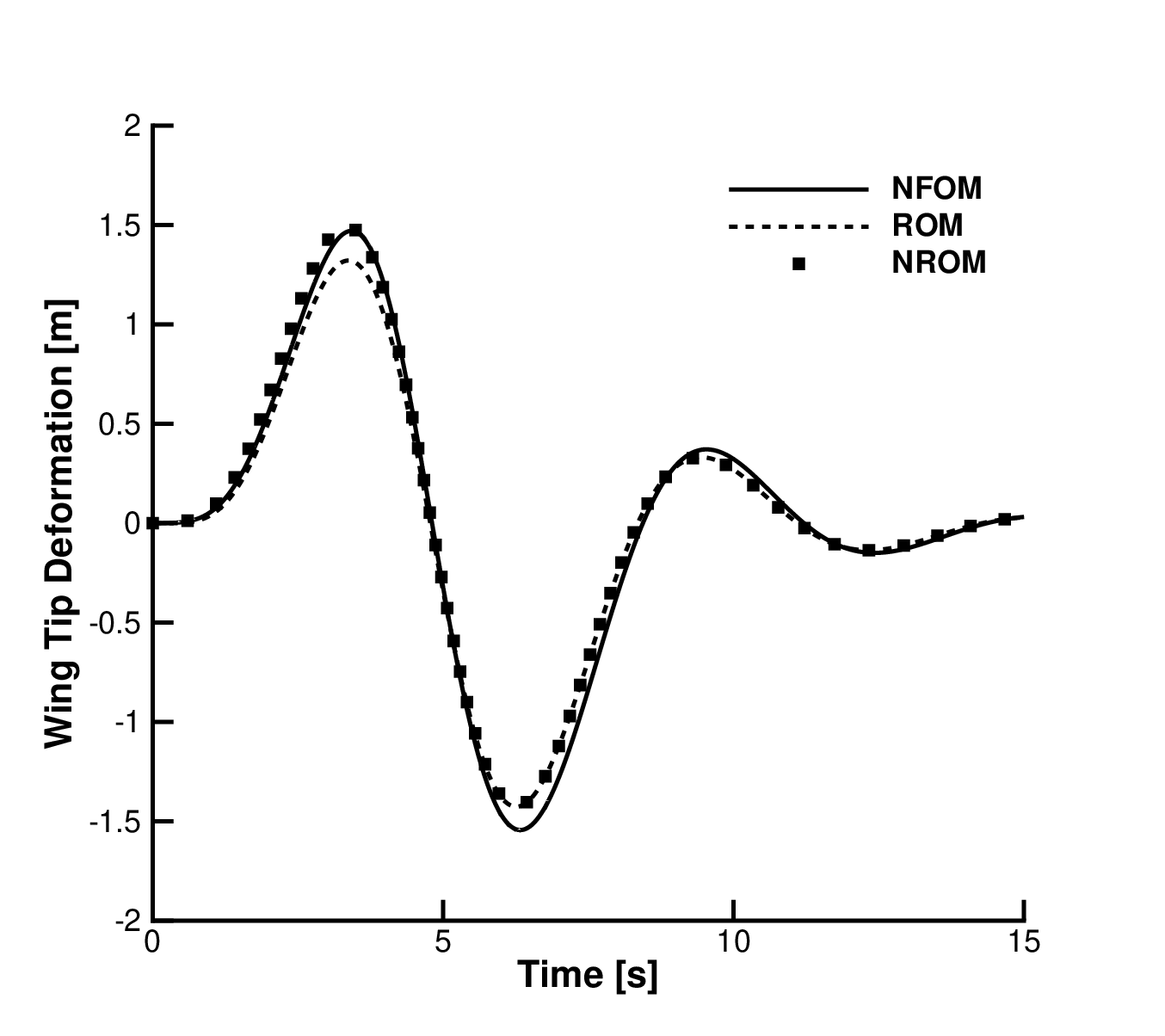}
\caption{Wing-tip vertical displacement}
\label{fig:vfa_wtip_nonlinear}
\end{subfigure}
\hfill
\begin{subfigure}[b]{0.48\textwidth}
\includegraphics[width=\textwidth]{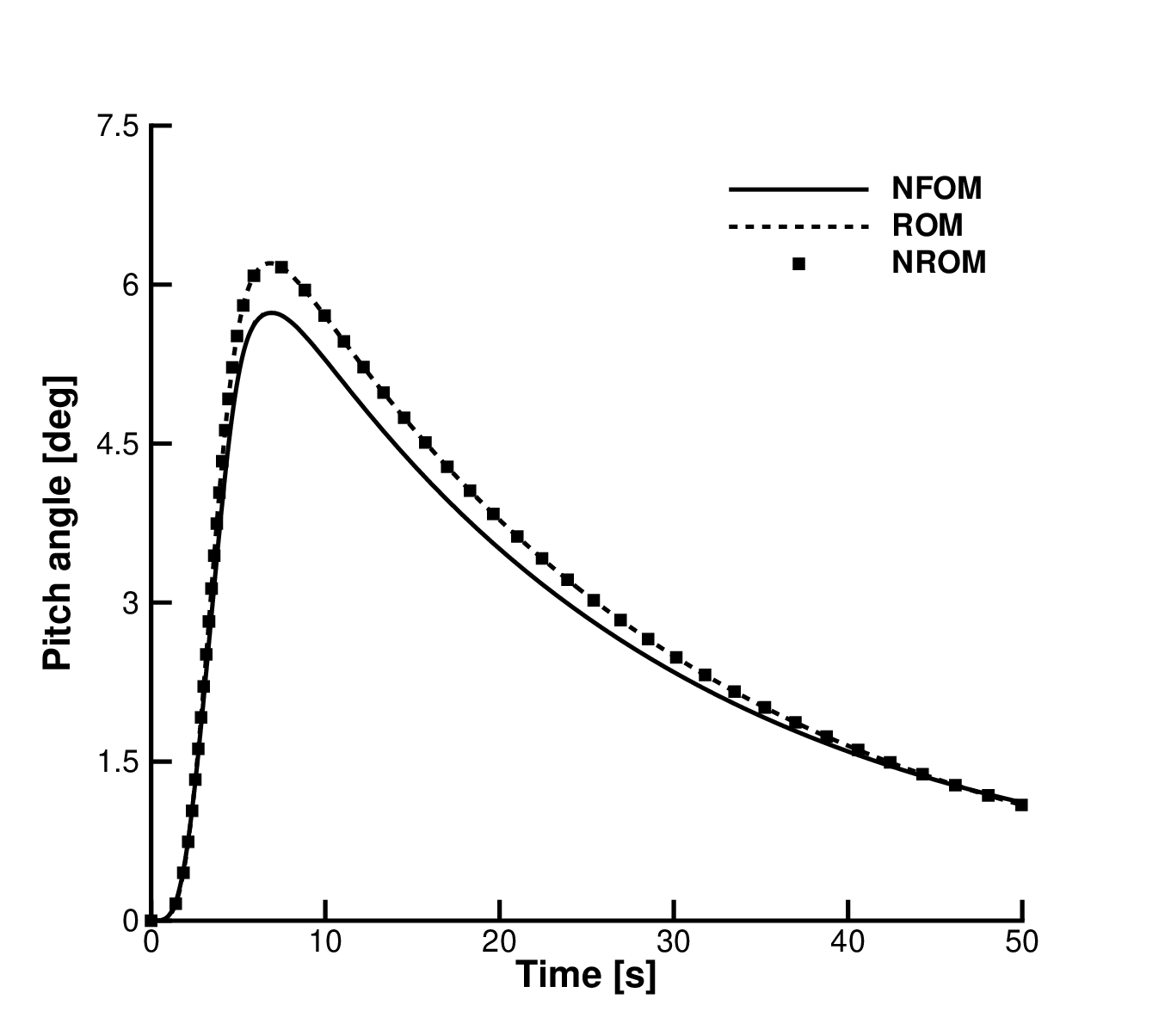}
\caption{Body angle of attack}
\label{fig:vfa_aoa_nonlinear}
\end{subfigure}
\caption{Time histories at the identified worst-case gust condition ($H_g = 125$~m): full-order nonlinear model (solid black), nonlinear ROM (dashed red), linear ROM (dotted blue). The linear ROM overestimates the response amplitude due to missing geometric stiffening.}
\label{fig:vfa_nonlinear}
\end{figure}

\Cref{fig:vfa_nonlinear} compares the time histories of wing-tip displacement and body angle of attack for the worst-case gust, computed with the full-order model, the nonlinear ROM, and the linear ROM.

The linear ROM accurately predicts the initial response while deformations remain small, but progressively overpredicts the amplitude as deformations grow and geometric stiffening effects become significant. At peak deformation, the linear ROM overpredicts the tip displacement due to the missing geometric stiffening. The nonlinear ROM with second-order Taylor expansion maintains accuracy throughout the entire response, confirming the sufficiency of quadratic nonlinear terms for capturing the geometric stiffening~\citep{Tantaroudas2015scitech, Tantaroudas2017bookchapter}.

\subsubsection{ROM convergence with number of modes}

\begin{figure}[htbp]
\centering
\includegraphics[width=0.6\textwidth]{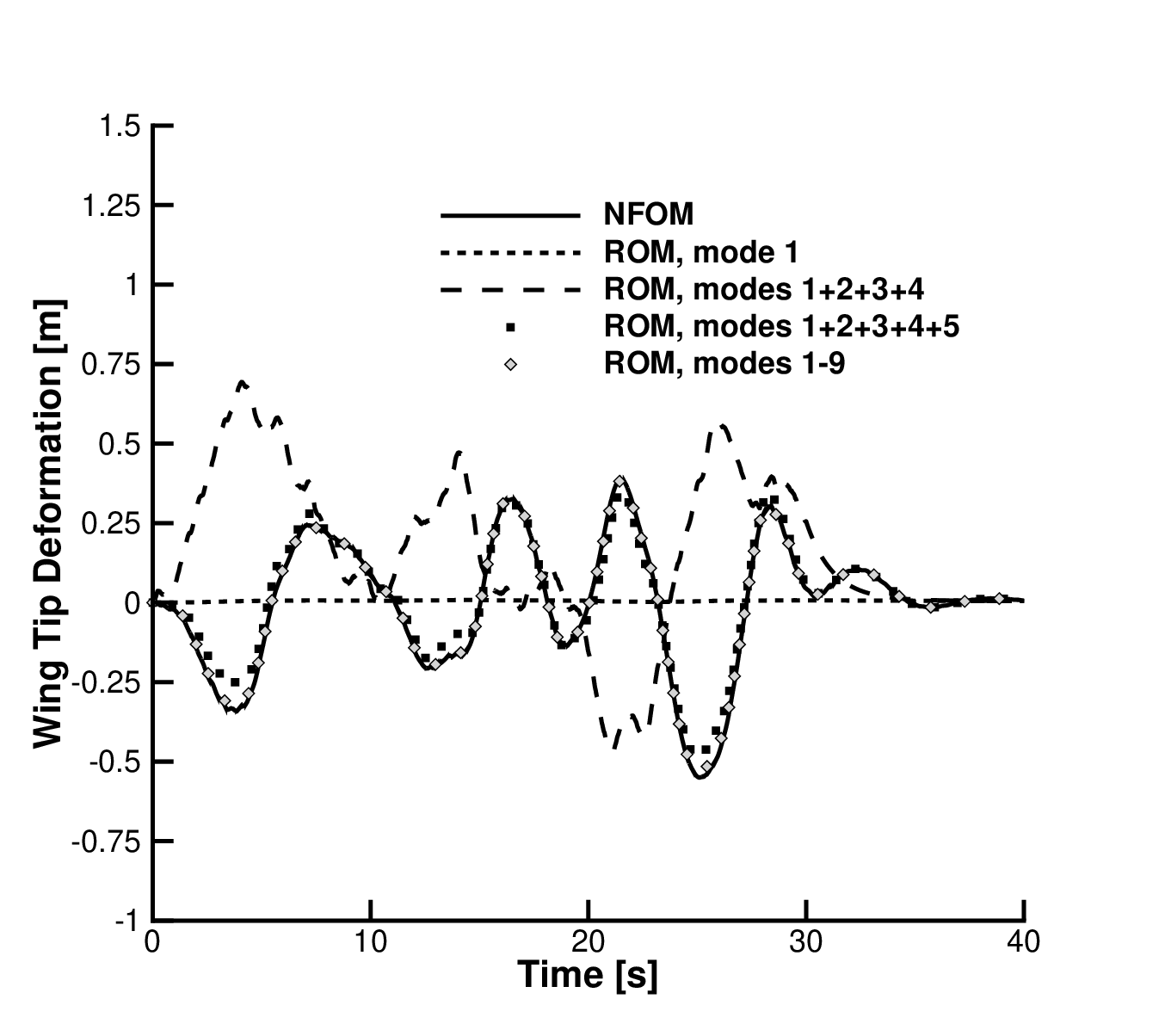}
\caption{Convergence of the ROM wing-tip response at the worst-case gust condition with increasing number of retained modes: 4 modes (rigid-body only), 5 modes (+1st bending), 7 modes (+2nd bending, 1st torsion), and 9 modes (converged). Full-order result shown for reference.}
\label{fig:vfa_convergence}
\end{figure}

\Cref{fig:vfa_convergence} demonstrates the convergence of the ROM worst-case gust response as modes are progressively added to the basis. Convergence of the wing-tip displacement is achieved with 9 modes; the rigid-body modes (1--4) capture the mean displacement, the first bending mode (5) dominates the oscillatory content, and higher modes (6--9) refine the high-frequency detail.

\subsubsection{Effect of structural flexibility on flight dynamics}

\begin{figure}[htbp]
\centering
\includegraphics[width=0.6\textwidth]{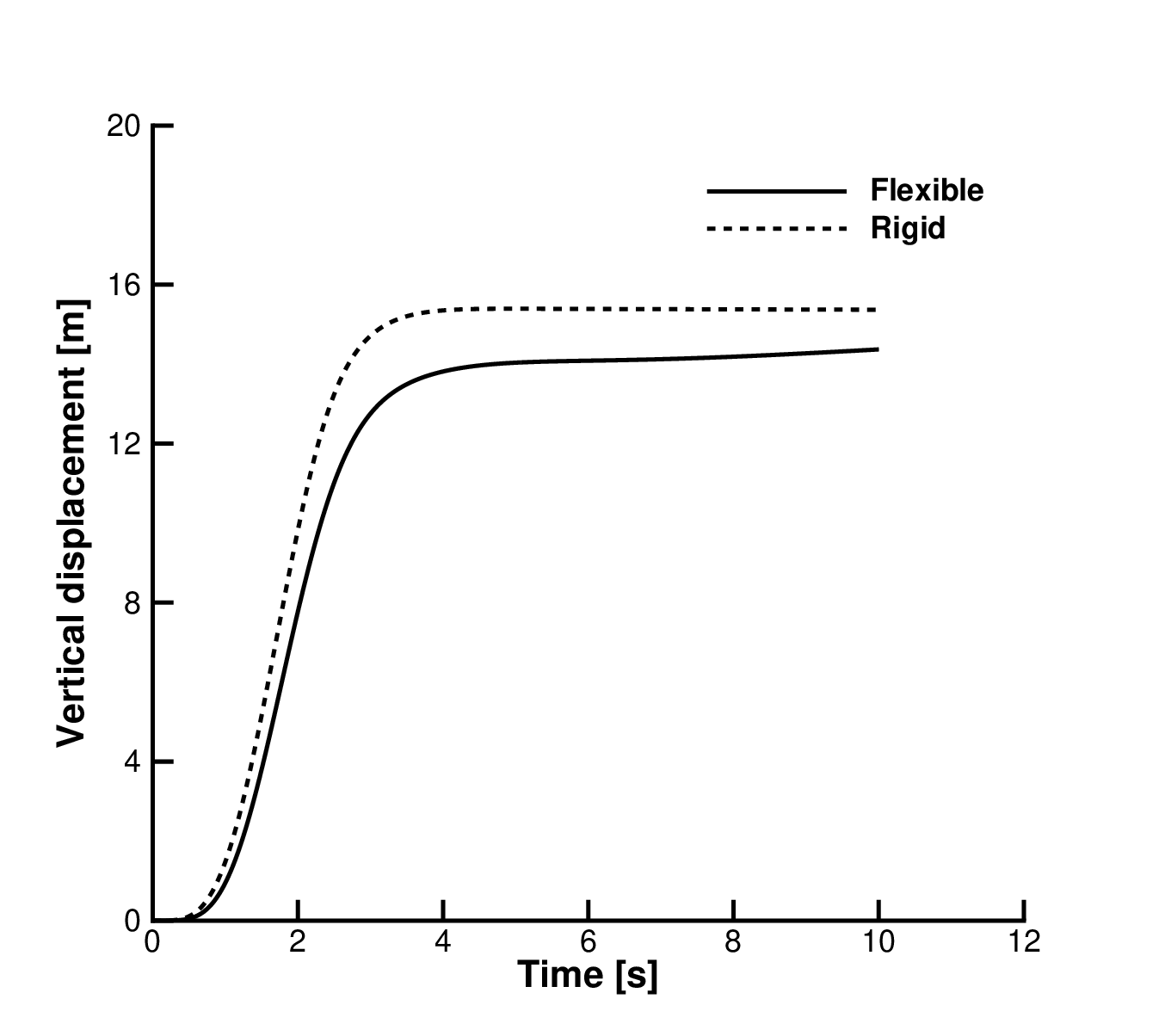}
\caption{Influence of structural flexibility on the vertical displacement response at the worst-case gust condition: comparison between the flexible aircraft model (solid) and the equivalent rigid aircraft (dashed). The deviation between the two trajectories demonstrates the importance of including aeroelastic coupling in gust response predictions for very flexible configurations.}
\label{fig:vfa_vert_comp}
\end{figure}

\Cref{fig:vfa_vert_comp} presents the effect of structural flexibility on the vertical displacement response at the worst-case gust condition. The flexible aircraft (solid) exhibits a measurably different trajectory from the rigid aircraft (dashed), confirming that flexibility effects significantly influence the flight dynamic response and must be accounted for in worst-case gust analysis.

\section{Discussion}
\label{sec:discussion}

\subsection{Practical Guidelines for ROM Selection}

Based on the results across three test cases, the following guidelines emerge for selecting the appropriate ROM fidelity level in worst-case gust searches. The linear ROM is adequate when maximum wing-tip deformations remain below approximately 10\% of the wingspan, corresponding to moderately flexible configurations or low-intensity gust scenarios; it provides the fastest computation and is suitable for initial screening. The nonlinear ROM with second-order Taylor expansion becomes essential when tip deformations exceed 10\% of the wingspan, as geometric stiffening significantly modifies the response and the linear ROM overpredicts loads in this regime, potentially leading to incorrect worst-case identification.

\subsection{Integration into Certification Workflows}

The ROM-based worst-case gust search can be integrated into certification workflows as a two-stage process~\citep{Tantaroudas2017bookchapter}:

\textbf{Stage 1 (Screening):} The ROM is used to sweep the full $(H_g, \text{flight condition})$ parameter space at negligible computational cost, identifying candidate worst-case conditions and ranking them by severity.

\textbf{Stage 2 (Validation):} A small number of full-order nonlinear simulations are performed at the candidate worst-case conditions to confirm the ROM predictions and provide high-fidelity design loads.

This approach significantly reduces the total computational effort compared to a brute-force full-order search, with speedups ranging from 30$\times$ (UAV configuration) to 600$\times$ (VFA flying-wing), while maintaining the fidelity of the final design loads.

\subsection{Limitations and Extensions}

The current framework employs strip-theory aerodynamics, which is adequate for preliminary design and high-aspect-ratio wings at moderate angles of attack. For detailed certification, upgrade to higher-fidelity methods (UVLM or panel method) may be necessary~\citep{Murua2012, DaRonch2014scitech_flight}. The NMOR framework is fully compatible with such upgrades, as it operates on the Jacobian eigenspectrum regardless of internal model structure~\citep{DaRonch2012rom}.

Extension to multi-dimensional gust fields (lateral gusts, gust gradients) and combined gust-manoeuvre scenarios is straightforward within the ROM framework, requiring only additional disturbance input channels in~\Cref{eq:rom}.

\section{Conclusions}
\label{sec:conclusions}

A reduced-order model based methodology for rapid worst-case gust identification has been presented and validated across three test cases of increasing complexity. The principal findings are:

\begin{enumerate}
\item Computational speedups of up to \textbf{600$\times$} are achieved for parametric worst-case gust searches, reducing total computation time from 222~hours to 22~minutes for a 1,616-DOF very flexible flying-wing.
\item The linear ROM provides sufficient accuracy for configurations with tip deformations below 10\% of the wingspan. For larger deformations, the nonlinear ROM with second-order Taylor expansion is essential to capture geometric stiffening.
\item The worst-case gust is not a monotonic function of the gust gradient distance: intermediate lengths produce the largest structural loads through resonant excitation. This non-trivial dependence necessitates exhaustive parametric searches, which are only computationally feasible with the ROM approach.
\item The methodology is directly applicable to certification workflows as a screening tool: the ROM rapidly identifies candidate worst-case conditions, validated by a small number of full-order simulations.
\item The framework is extensible to higher-fidelity aerodynamic models (CFD, UVLM) without modification of the reduction procedure, as it operates on the Jacobian eigenspectrum of the coupled system~\citep{DaRonch2012rom, Tantaroudas2017bookchapter}.
\end{enumerate}

\section*{Acknowledgements}

This work was supported by the U.K.\ Engineering and Physical Sciences Research Council (EPSRC) grant EP/I014594/1 on ``Nonlinear Flexibility Effects on Flight Dynamics and Control of Next-Generation Aircraft.'' The authors are grateful to Prof.\ A.\ Da Ronch and Prof.\ K.J.\ Badcock for their valuable guidance on flight dynamics modelling.

\bibliographystyle{unsrtnat}
\bibliography{references}

\end{document}